\documentclass{iopart}
\usepackage{iopams}
\usepackage{setstack}
\usepackage{url}
\usepackage{graphicx}	% figures
\begin{document}

%\title{Visualizing the edge of locality: the appearance of the horizon from inside a black hole}
\title{The edge of locality: visualizing a black hole from the inside}

\author{Andrew J S Hamilton$^{1,2}$ and Gavin Polhemus$^{1,3}$}
\address{$^1$JILA, Box 440, U. Colorado, Boulder, CO 80309, USA}
\address{$^2$Dept.\ Astrophysical \& Planetary Sciences, Box 391, U. Colorado, Boulder, CO 80309, USA}
\address{$^3$Poudre High School, 201 Impala Drive, Fort Collins, CO 80521, USA}
\eads{\mailto{Andrew.Hamilton@colorado.edu}, \mailto{gavinpolhemus@comcast.net}}
%\author{Colin S Wallace}
%\ead{Colin.Wallace@colorado.edu}
%\address{Dept.\ Astrophysical \& Planetary Sciences,
%Box 391, U. Colorado, Boulder CO 80309, USA}

\newcommand{\simlt}{\alt}
\newcommand{\simgt}{\agt}

\newcommand{\dd}{\rmd}
\newcommand{\DD}{D}
\newcommand{\im}{\rmi}
\newcommand{\ee}{\rme}

\newcommand{\emit}{{\rm em}}
\newcommand{\obs}{{\rm obs}}

\newcommand{\bJ}{\bi{J}}
\newcommand{\bL}{\bi{L}}

\newcommand{\unit}[1]{\, {\rm#1}}

%\hyphenpenalty=3000

%--------------------
% FIG
\newcommand{\coordfig}{
    \begin{figure}
    \begin{center}
    \leavevmode
    \includegraphics[scale=.7]{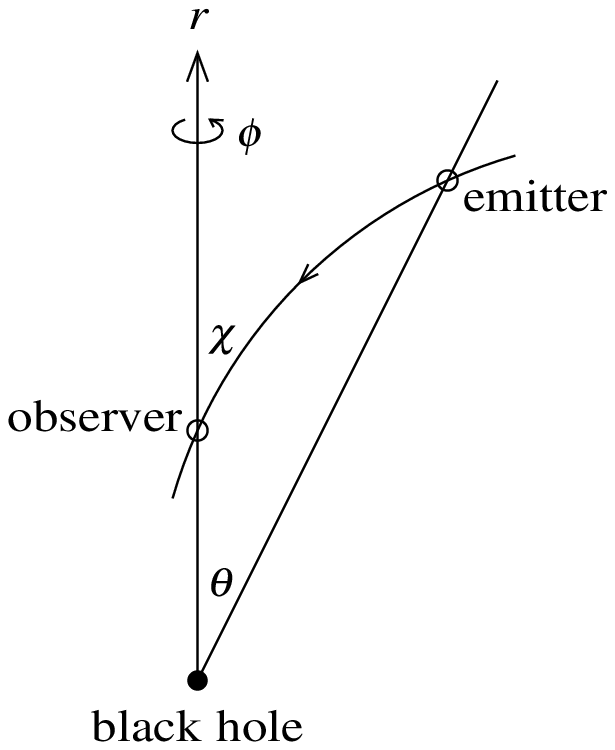}
    \caption[1]{
    \label{coord}
Coordinate system.
The emitter and observer are separated by angle $\theta$
in a polar coordinate system
$\{ r , \theta, \phi \}$
about the black hole.
A light ray from emitter to observer
subtends apparent angle
$\chi$
from the vertical axis.
%from the black hole to observer.
    }
    \end{center}
    \end{figure}
}

%--------------------
% FIG
\newcommand{\bhholoschwfig}{
    \begin{figure}[b]
    \begin{center}
    \leavevmode
    \includegraphics[scale=1]{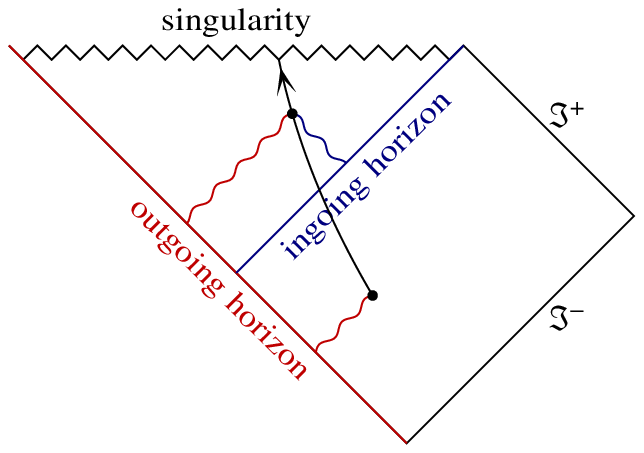}
    \caption[1]{
    \label{bhholo_schw}
Penrose diagram of a Schwarzschild black hole.
The arrowed line represents the worldline of an infaller.
The wiggly lines represent outgoing and ingoing photons
emitted from the outgoing and ingoing horizons.
    }
    \end{center}
    \end{figure}
}

%--------------------
% FIG
\newcommand{\bhholopenrosefig}{
    \begin{figure}
    \begin{center}
    \leavevmode
    \includegraphics[scale=1]{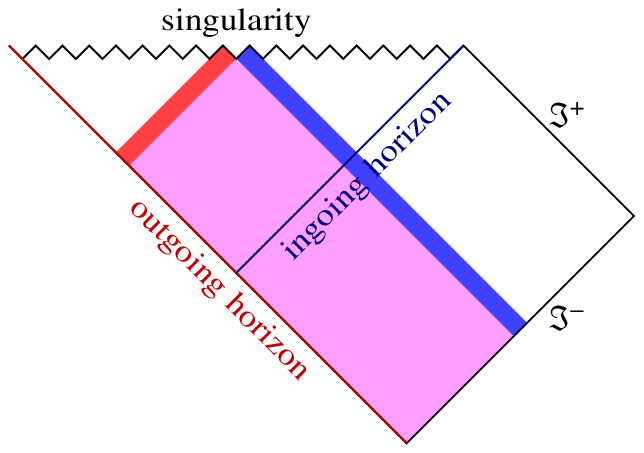}
    \caption[1]{
    \label{bhholo_penrose}
(Color online)
Causal diamond (mauve) of an observer who hits the singularity
of a Schwarzschild black hole.
Null rays in opposite directions
(thick red and blue lines)
along the past lightcone of such an observer
lie at the edge of locality,
the future lightcones of such points being on the cusp of intersecting.
Observer complementarity asserts that
locality applies within the causal diamond of any observer,
but breaks down between spacelike-separated points
that lie outside any observer's causal diamond,
that is, whose future lightcones do not intersect.
    }
    \end{center}
    \end{figure}
}

%--------------------
% FIG
\newcommand{\diamondfig}{
    \begin{figure}
    \leavevmode
    \begin{center}
    \includegraphics[scale=1]{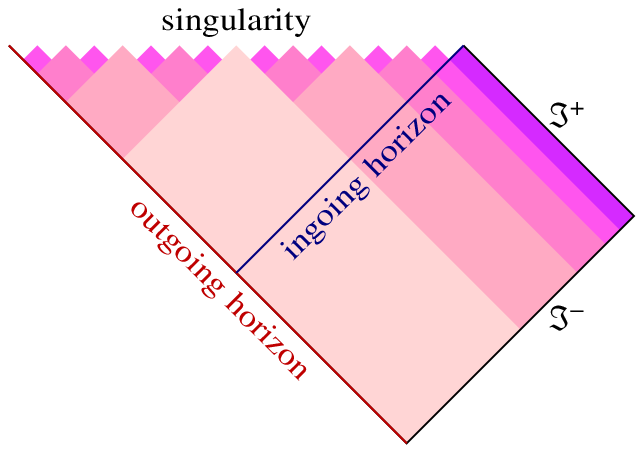}
    \caption[1]{
    \label{bhholo_diamond}
(Color online)
The number of regions
with non-intersecting future lightcones
proliferates near the singularity of a Schwarzschild black hole.
    }
    \end{center}
    \end{figure}
}

%--------------------
% FIG
\newcommand{\sceneonefig}{
    \begin{figure}
    \begin{center}
    \leavevmode
    \includegraphics[scale=.95]{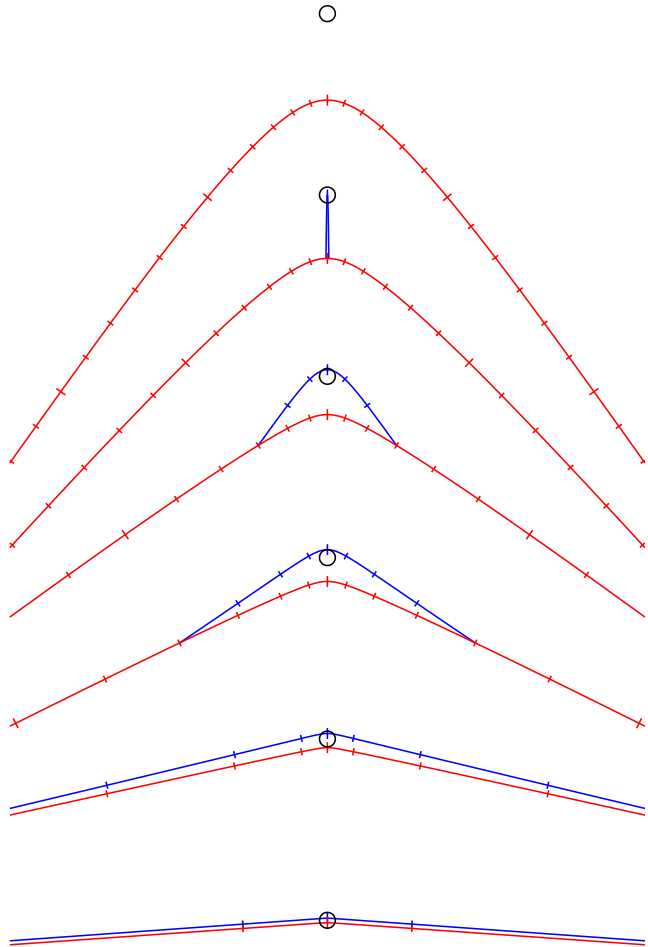}
    \caption[1]{
    \label{scene1}
(Color online)
Six successive views of the
location of the ingoing (blue) and outgoing (red) horizons
of a Schwarzschild black hole
as perceived by an observer
who is free-falling radially from zero velocity at infinity.
From top to bottom,
the observer
is at radii of
$3$, $1.999$, $1$, $0.5$, $0.1$, and $0.01$ geometric units.
The small circle, diameter $1$ geometric unit,
at the center of each frame
marks the position of the observer.
Tick marks on the horizons are spaced every $30^\circ$.
    }
    \end{center}
    \end{figure}
}

%--------------------
% FIG
\newcommand{\sceneobsonefig}{
    \begin{figure}
    \begin{center}
    \leavevmode
    \includegraphics[scale=.95]{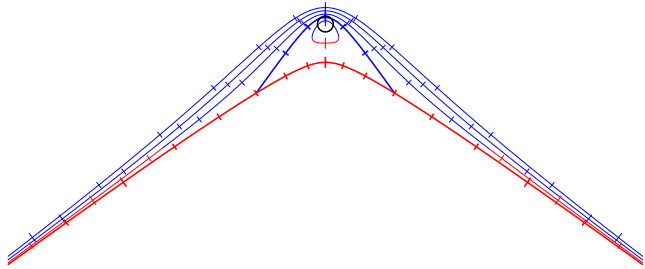}
    \caption[1]{
    \label{sceneobs1}
(Color online)
Additional detail for the observer at $1$ geometric unit,
in the third-from-top frame of Figure~\protect\ref{scene1}.
The various lines show the perceived location of emitting surfaces at,
from top to bottom,
radii of
$3.5$, $3$ (photon sphere), $2.5$, $2$ (horizon), and $1.5$ geometric units.
The lines are blue where the emitted photon is initially ingoing,
and red where the emitted photon is initially outgoing.
    }
    \end{center}
    \end{figure}
}

%--------------------
% FIG
\newcommand{\scenefig}{
    \begin{figure}
    \begin{center}
    \leavevmode
    \includegraphics[scale=.9]{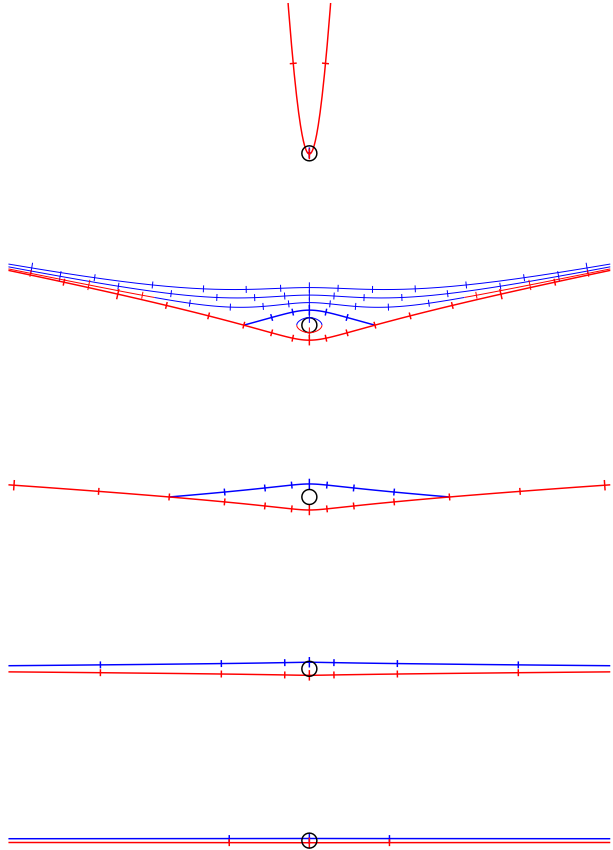}
    \caption[1]{
    \label{scene}
(Color online)
Similar to Figure~\protect\ref{scene1},
but as perceived by an observer
who is infalling on the zero-energy radial geodesic,
which makes manifest the symmetry between ingoing and outgoing.
The zero-energy geodesic exists only inside the horizon,
so only images where the observer is inside the horizon can be shown.
From top to bottom,
the observer is at radii of
$1.999$, $1$, $0.1$, $0.5$, and $0.01$ geometric units,
as in all but the top frame of Figure~\protect\ref{scene1}.
The second-from-top frame,
where the observer is at $1$ geometric unit,
shows additional detail
similar to Figure~\ref{sceneobs1};
the various lines show the perceived location of emitting surfaces at,
from top to bottom,
radii of
$3.5$, $3$ (photon sphere), $2.5$, $2$ (horizon), and $1.5$ geometric units.
The lines are blue where the emitted photon is initially ingoing,
and red where the emitted photon is initially outgoing.
    }
    \end{center}
    \end{figure}
}

%--------------------
% FIG
\newcommand{\zoomfig}{
    \begin{figure}
    \begin{center}
    \leavevmode
    \includegraphics[scale=.9]{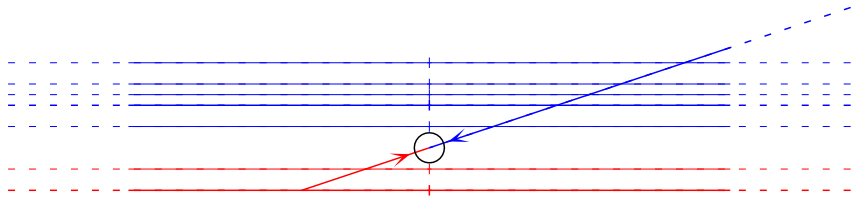}
    \caption[1]{
    \label{zoom}
(Color online)
Zoom-in of the view perceived by an observer
close to the singularity,
at a radius of $10^{-8}$ geometric units.
The central circle, diameter
$10^{-4}$ geometric units,
marks the position of the observer,
who is infalling on the zero-energy radial geodesic,
as in Figure~\ref{scene}.
The horizontal lines represent perceived locations of surfaces at,
starting farthest from the observer,
radii of
$4$, $3$ (photon sphere), $2.5$, $2$ (horizon), and $1$ geometric units.
The lines are blue where the emitted photon is ingoing,
red where the emitted photon is outgoing.
The dashed lines are a reminder that the lines extend horizontally
way beyond shown here.
%but this close to the singularity the view
%is much the same regardless of the radial motion of the observer,
%unless the observer is accelerating like crazy.
The arrowed lines
converging on the observer
represents a sample pair of
oppositely directed ingoing and outgoing light rays
along the observer's past lightcone.
Any pair of points on the oppositely directed light rays
(one ingoing, one outgoing)
are near the edge of locality;
that is, their future lightcones intersect
just barely before the singularity.
    }
    \end{center}
    \end{figure}
}

%--------------------
% FIG
\newcommand{\blueshiftfig}{
    \begin{figure}
    \begin{center}
    \leavevmode
    \includegraphics[scale=.7]{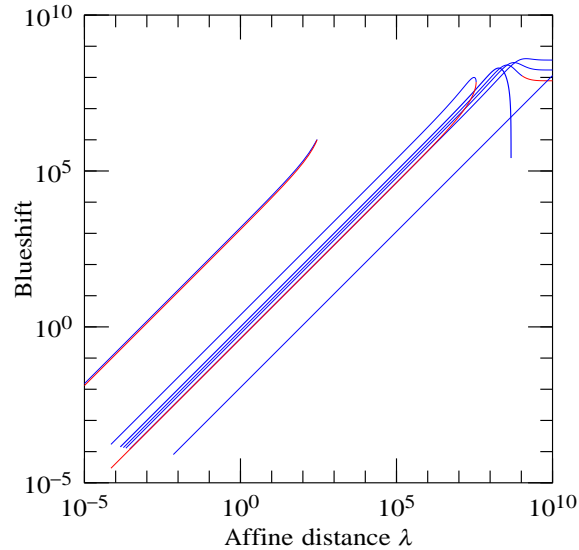}
    \caption[1]{
    \label{blueshift}
(Color online)
Blueshift (= ratio of observed to emitted energy)
observed by an observer
close to the singularity,
at a radius of $10^{-8}$ geometric units,
the same as in Figure~\protect\ref{zoom}.
The blueshift is a plotted as a function of the affine distance
between the observer and the emitting point.
The lines represent emitting surfaces at,
from top to bottom,
radii of
$0.01$,
$1$,
$2$ (horizon),
$2.5$,
$3$ (photon sphere),
$4$,
and $100$
geometric units.
The lines are blue where the emitted photon is ingoing,
red where the emitted photon is outgoing.
    }
    \end{center}
    \end{figure}
}

%--------------------
% FIG
\newcommand{\distfig}{
    \begin{figure}
    \begin{center}
    \leavevmode
    \includegraphics[scale=.7]{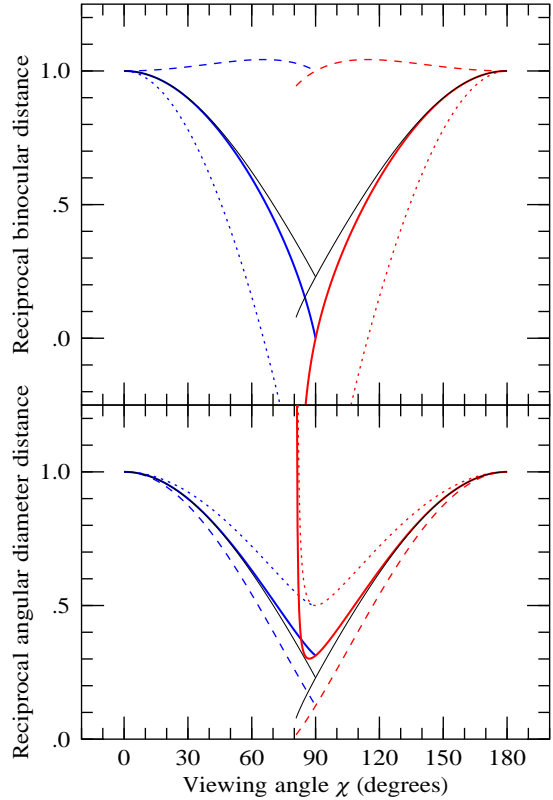}
    \caption[1]{
    \label{dist}
(Color online)
Perceived
reciprocal
binocular
(upper panel)
and
angular diameter
(lower panel)
distances
to the ingoing (blue) and outgoing (red) horizons
of a Schwarzschild black hole
perceived by an observer inside the black hole,
as a function of the viewing angle $\chi$ relative to the vertical axis,
with $0^\circ$ being up to the sky above,
and $180^\circ$ down to the black hole below.
The observer
is at a radius of 1 geometric unit,
and
is infalling on the zero-energy radial geodesic,
the same as illustrated in the top image of Figure~\protect\ref{scene}.
Dashed and dotted lines show the reciprocal distances
respectively in the polar and azimuthal directions,
while thick solid lines show the average of the polar and azimuthal
reciprocal distances.
The averaged reciprocal distances
provide a good estimate of the affine distance
(thin solid black lines)
for viewing angles not too far off axis.
%On axis (at $0^\circ$ and $180^\circ$)
%the perceived distance to the horizon at 2 geometric units
%is 1 geometric unit by all measures.
    }
    \end{center}
    \end{figure}
}

%--------------------
% FIG
\newcommand{\apefig}{
    \begin{figure}
    \begin{center}
    \leavevmode
    \includegraphics[scale=.5]{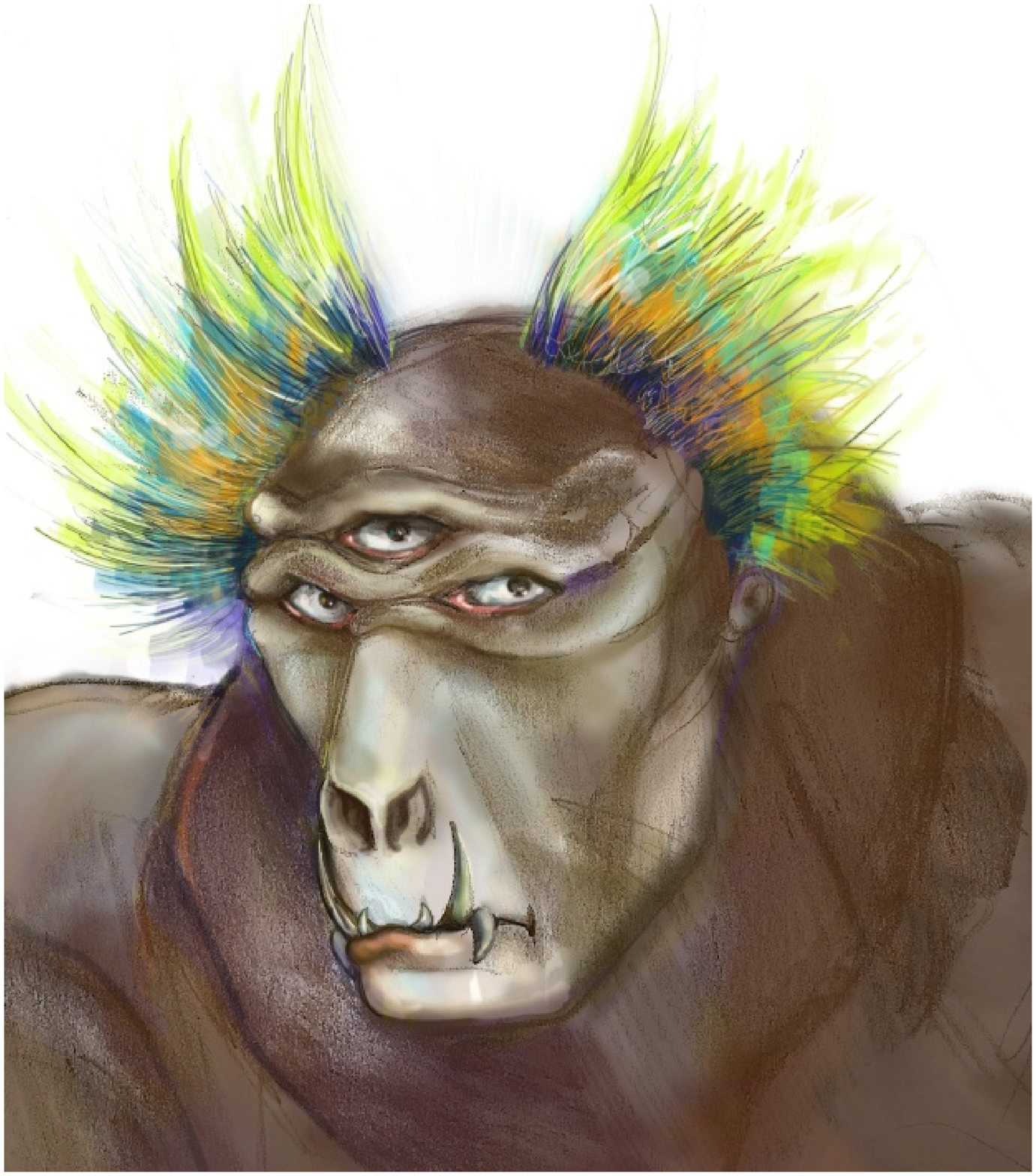}
    \caption[1]{
    \label{ape}
A three-eyed inhabitant of a highly curved spacetime.
    }
    \end{center}
    \end{figure}
}

\begin{abstract}
We illustrate and discuss the view seen by an observer
inside the horizon of a Schwarzschild black hole.
The view as the observer approaches the central singularity
is of particular interest
because,
according to ideas arising from ``observer complementarity,''
points in opposite directions
along the observer's past lightcone
are at ``the edge of locality,''
where standard flat-space quantum-field-theory commutation rules may
be at the brink of failure.
Near the singularity,
the observer's view is aberrated by the diverging tidal force
into a horizontal plane.
The view in the horizontal plane is highly blueshifted,
but all directions other than horizontal appear highly redshifted.
We argue that the affine distance provides a canonical
measure of distance along a light ray from emitter to observer.
Since the affine distance is not directly measurable by the observer,
we also consider perceptual distances,
and argue that
the trinocular distance
(binocular distance is inadequate)
provides an estimate of affine distance
that would allow tree-leaping apes
to survive in highly curved spacetime.
\end{abstract}

	% Classical general relativity

\date{\today}

\maketitle

\section{Introduction}

What does it look like when you fall inside the horizon
of a (Schwarzschild, for simplicity) black hole?
Based on your experience in the 3D world,
you might imagine that falling through the horizon
would be like falling through any other surface.
But it is not.
%In their guide to the literature on black holes,
%\cite{GM09}
%describe the general relativistic animations at
%\cite{H98}
%as have
%``a somewhat confusing portrayal of the horizon.''
%The reason it is confusing to \cite{GM09} is that the actual appearance
%does not conform to their preconceptions.

\bhholoschwfig

As its Penrose diagram shows,
Figure~\ref{bhholo_schw},
the Schwarzschild geometry contains not one but two distinct horizons,
the future horizon, and the past horizon.
The future horizon may be called the ``ingoing'' horizon,
since photons that fall through it are necessarily ingoing,
falling towards the black hole.
The past horizon may be called the ``outgoing'' horizon,
since photons that emanate from it
are either moving away from the black hole (outside the horizon),
or trying (and failing) to do so (inside the horizon).

Of course,
in a real black hole formed from the collapse of the core of a star,
there is no past horizon.
The past horizon is replaced by an exponentially redshifting
region that contains the collapsing star and its interior
(and beyond).
However, as time goes by,
the exponentially redshifting image of the collapsing star
becomes more and more indistinguishable from the outgoing horizon
of a Schwarzschild black hole.
In this paper,
we suppose that the black hole collapsed long enough ago
that it has become effectively indistinguishable
from a Schwarzschild black hole,
and we refer to the exponentially redshifting surface
as the outgoing horizon.

As
Figure~\ref{bhholo_schw} shows,
when an observer outside the horizon observes the horizon of a black hole,
they are actually observing the outgoing horizon.
When they subsequently fall through the horizon,
they do not fall through the horizon they were looking at, the outgoing horizon;
rather, they fall through the ingoing horizon, which was invisible
to them until they actually passed through it.
Once inside the horizon,
the infaller sees both outgoing and ingoing horizons.
The outgoing horizon
appears ahead of the infaller,
in the direction towards the black hole,
just as it did when the infaller was outside the horizon.
The ingoing horizon
appears behind the infaller,
in the direction away from the black hole.
Together, the outgoing and ingoing horizons
form apparently 2D surfaces that encompass the infaller,
the ``Schwarzschild bubble.''
The outgoing and ingoing horizons meet at a circle;
at this circle,
the angular momentum per unit energy of emitted photons is infinite.
%(in the Schwarzschild limit).

In case the reader is suspicious that the Penrose diagram,
Figure~\ref{bhholo_schw},
may not capture the scene reliably,
because the diagram is conformal,
the reader is invited to view the general relativistically
ray-traced visualizations at
\cite{Hamilton:2009}.
The visualizations confirm that an infaller
sees distinct outgoing and ingoing horizons,
consistent with the Penrose diagram.

The question of
what things look like when you fall
inside the horizon of a black hole is of some intrinsic interest.
Certainly public audiences are fascinated by it.

But there are deeper reasons, having to do with quantum gravity
and the breakdown of locality across horizons,
for being curious about what things look like inside a black hole.
Locality is the standard flat-space quantum-field-theory proposition
that operators at spacelike-separated points commute.
By the principle of equivalence,
one might expect that locality would hold in spacetimes
where the curvature is significantly below Planck,
which includes the inside of a black hole except near its singularity.
However, if locality holds,
then unitarity must break down in black hole evaporation,
because points inside the black hole are spacelike-separated from points
outside the black hole after it has evaporated,
so information in points inside the black hole is lost to the outside
\cite{Hawking:1976ra}.
Unitarity is widely considered to be a more fundamental principle than locality.
If so, then it must be that locality breaks down across the
horizon of a black hole.
The interior of a black hole must share states with the exterior,
an idea called ``black hole complementarity''
\cite{Susskind:1993if}.

\bhholopenrosefig

\diamondfig

Black hole complementarity has led to a more general concept
known as ``observer complementarity''
\cite{Parikh:2002py,Dyson:2002pf}.
Observer complementarity posits that locality breaks down
between any two spacelike-separated points
that lie outside the causal diamond of any observer.
That is, locality breaks down between any two points
whose future lightcones do not intersect.
Figure~\ref{bhholo_penrose}
illustrates the idea of observer complementarity.
It shows the causal diamond of an observer
who hits the central singularity of a Schwarzschild black hole.
%The past lightcone of such an observer constitutes the observer's horizon.
Observer complementarity implies that locality applies within
the observer's causal diamond,
but breaks down outside it,
so that points within the causal diamond
share states with those outside.
In its strongest form,
observer complementarity asserts that the causal diamond of the observer
contains a complete set of states.

Recently,
we \cite{Wallace:2008zz}
showed that classical processes of dissipation
can create many orders of magnitude more than the Bekenstein-Hawking
\cite{Bekenstein:1973ur,Hawking:1976de}
entropy inside the horizon of a black hole.
This would lead to a gross violation of the second law of thermodynamics
if the black hole subsequently evaporated radiating only the
Bekenstein-Hawking entropy.
The calculation in \cite{Wallace:2008zz} made the classical assumption
that entropy is additive over spacelike slices inside the black hole,
which is valid as long as locality holds inside the black hole.
The second law of thermodynamics can be rescued
provided that there is a wholesale breakdown of locality inside the black hole.
Observer complementarity predicts precisely such a wholesale breakdown,
since, as illustrated in
Figure~\ref{bhholo_diamond},
the number of regions with non-intersecting future lightcones
proliferates near the singularity of a black hole.
Detailed calculations
\cite{Polhemus:2009vv}
confirm that, if observer complementarity holds,
then the calculation of \cite{Wallace:2008zz}
does not lead to a violation of the second law.
This is cogent evidence in favor of observer complementarity.

The question of exactly how locality breaks down remains a subject of debate
\cite{Giddings:2006sj,Giddings:2007pj,Hogan:2007ci,Hogan:2008ir}.
The present paper does not consider the question per se.
Rather, it addresses what things look like classically
in the regime where locality is on the threshold of breaking down.
Specifically,
as illustrated in
Figure~\ref{bhholo_penrose},
points in opposite directions along the past lightcone
of an observer at the instant of hitting the singularity
lie at ``the edge of locality,''
the future lightcones of such points just barely intersecting
as they hit the singularity.

Regardless of motivation,
the purpose of the present paper
is to describe what things look like from the perspective
of an observer inside the horizon of a Schwarzschild black hole,
especially as they approach the central singularity,
where, from the observer's point of view,
the wavefunction of the universe builds to its ultimate climax.

An important technical issue addressed in this paper
is the problem of assigning a perceived location to an emitting object.
An observer in a highly curved spacetime
has no difficulty in specifying precisely the perceived
angular position of an object,
but has a harder time determining its distance.
In this paper we argue that the affine distance
provides a definitive measure of distance along
null rays in curved spacetime.
However, the affine distance is not always directly measurable,
and we therefore pose the question of whether there are
direct measures of distance that adequately approximate
the affine distance.
We argue that the trinocular distance
(binocular vision is inadequate)
provides a viable estimate of the affine distance.

The paper is structured as follows.
Section~\ref{affine}
discusses the affine distance.
The core of the paper is
\S\ref{sceneinside},
which discusses the scene inside the horizon.
Finally,
\S\ref{perceptual}
discusses perceptual surrogates for the affine distance.

\section{Affine distance}
\label{affine}

\coordfig

%Reality is what is seen,
%but what does that mean?

The canonical measure of distance along a null ray in general relativity
is the affine distance $\lambda$,
which is the proper distance along the null ray measured by an ensemble of
observers arrayed along it each of whom measures
the ray to have the same photon frequency.
The arbitrary overall constant in the definition of affine distance
is fixed naturally by tying it to the reference frame of the observer.
If you lived in a highly curved spacetime,
and you put out your rigid arm to touch something,
then the distance that your arm would measure would be the affine distance
(see \ref{rigidarm} for clarification of this assertion).

The Schwarzschild metric is,
in polar coordinates
$x^\mu \equiv \{ t , r , \theta , \phi \}$,
\begin{equation}
  \dd s^2
  =
  - \, B \, \dd t^2
  + B^{-1} \dd r^2
  + r^2 ( \dd \theta^2 + \sin^2\!\theta \, \dd \phi^2 )
  \ ,
\end{equation}
where
$B \equiv 1 - 2 M / r$
with $M$ the mass of the black hole.
In the Schwarzschild metric,
the coordinate 4-velocity
$k^\mu \equiv \dd x^\mu / \dd \lambda$
along a null geodesic satisfies three conservation laws,
associated with
energy,
mass,
and
angular momentum $\bJ$,
\begin{equation}
  k^t = {1 / B}
  \ , \quad
  k^r = \pm \sqrt{1 - B J^2 / r^2}
  \ , \quad
  %k^\theta = {J / r^2}
  k^\perp = {\bJ / r^2}
  \ .
\end{equation}
Here
the photon energy $k^t$ has
been normalized to one as perceived by observers at rest at infinity.
Integrating $\dd r / \dd \lambda = k^r$ yields
the affine distance $\lambda$
normalized
to observers at rest at infinity:
\begin{equation}
  \lambda
  =
  \int_{r_\emit}^{r_\obs}
  {\dd r \over \sqrt{1 - B J^2 / r^2}}
  \ .
\end{equation}
The observed affine distance is then
\begin{equation}
  \lambda_\obs
  =
  E_\obs \lambda
\end{equation}
where
$E_\obs$
is the photon energy perceived by the observer
relative to the photon energy perceived by observers at rest at infinity.
If the observer has specific energy $E$ and angular momentum $\bL$,
then their coordinate 4-veolocity $u^\mu \equiv \dd x^\mu / \dd \tau$ is
\begin{equation}
  u^t = {E \over B}
  \ , \quad
  u^r = \pm \sqrt{E^2 - B ( 1 + L^2 / r^2 )}
  \ , \quad
  u^\perp = {\bL / r^2}
  \ .
\end{equation}
The observed photon energy
$E_\obs \equiv - u_\mu k^\mu$
is then
\begin{equation}
\label{Eobs}
  E_\obs
  =
  {E \mp \sqrt{( 1 - B J^2 / r^2 ) \left[ E^2 - B ( 1 + L^2 / r^2 ) \right]}
  \over B}
  -
  {\bL . \bJ \over r^2}
  \ .
\end{equation}

The view from inside a black hole is most symmetrical
in the frame of an observer radially free-falling
on the zero-energy geodesic, $E = 0$, $\bL = 0$,
at the border between ingoing ($E > 0$)
and outgoing ($E < 0$).
An observer on the zero-energy radial geodesic sees
a photon of angular momentum $J$
subtend an angle $\chi$ away from the vertical axis
(see Figure~\ref{coord}) given by
\begin{equation}
\label{J0}
  | \tan \chi |
  =
  \sqrt{- B}
  J / r
  \ .
\end{equation}
The zero-energy radial observer sees
the observed photon energy $E_\obs$, equation~(\ref{Eobs}),
relative to its energy at rest at infinity
to be
\begin{equation}
\label{Eobs0}
  E_\obs
  =
  \sqrt{ {1 - B J^2 / r^2 \over - B} }
  =
  {| \sec\chi | \over \sqrt{- B}}
  \ .
\end{equation}

\sceneonefig

\section{The scene inside the horizon}
\label{sceneinside}

\sceneobsonefig

The scene that any observer sees is, of course,
the three-dimensional hypersurface
that constitutes the past lightcone of the observer.

Figure~\ref{scene1}
shows (from top to bottom) six successive views of
the location of the ingoing and outgoing horizons
of a Schwarzschild black hole,
as perceived by an observer
who free-falls radially from zero velocity at infinity
($E = 1$, $\bL = 0$).
In these views,
the distance from any point on the scene to the observer
has been set equal to the affine distance.

In the top panel of Figure~\ref{scene1},
the observer is outside the horizon,
at a radius of $3$ geometric units
(which also happens to be the radius of the photon sphere).
Being outside the horizon,
the observer sees only the outgoing horizon,
not the ingoing horizon.
The outgoing horizon appears to recede into the infinite distance
(extending beyond the view shown in Figure~\ref{scene1}),
as photons emanating from the horizon are able to orbit
the black hole an infinite number of times near the photon sphere
before peeling off into the observer's eye.

In the second panel from top of Figure~\ref{scene1},
the observer is just inside the horizon,
at a radius of $1.999$ geometric units.
At the instant that the infaller passes through the ingoing horizon,
the ingoing horizon appears not as a surface,
but rather as a line (which appears in projection as a point)
that extends from the position of the observer down to the outgoing horizon.

In the third and fourth panels from top of Figure~\ref{scene1},
the observer is inside the horizon,
at radii of respectively $1$ and $0.5$ geometric units.
The ingoing horizon appears to form a bubble,
the ``Schwarzschild bubble,'' over the observer's head.
The ingoing (blue) and outgoing (red) horizons join at a circle
(projecting to a pair of points in Figure~\ref{scene1}),
where the photon angular momentum $J$ is infinite.
As the observer falls inward,
the Schwarzschild bubble expands horizontally.

In the limit as the observer approaches the singularity,
the edge of the Schwarzschild bubble extends to an angular radius of $\pi$
around the black hole,
so that the bubble finally encompasses
the full surface of each of the ingoing and outgoing horizons.

Figure~\ref{sceneobs1}
shows additional detail for the observer at a radius of $1$ geometric unit
(third frame from top in Figure~\ref{scene1}).
The Figure shows
the perceived location
of emitting surfaces at radii of
$3.5$, $3$, $2.5$, and $1.5$ geometric units.
Inside the horizon,
an observer can see points only at radii larger than their own:
points at smaller radii are invisible.
The observer sees
the surface at $1.5$ geometric units
to surround them.
Ingoing photons (blue) appear above the observer
(away from the black hole),
while outgoing photons (red) appear ahead of the observer
(towards the black hole).

As the observer approaches the singularity,
the horizons appear to flatten out,
and appear to approach each other ever more closely,
as illustrated in
the bottom two panels of Figure~\ref{scene1}.
The appearance is caused by the growing tidal field
near the singularity,
which aberrates the view so as to shift the apparent positions
of objects away from vertical towards horizontal.

\scenefig

Figure~\ref{scene}
shows essentially the same set of views as Figure~\ref{scene1},
but now from the perspective of an observer
free-falling radially on the zero-energy geodesic
($E = 0$, $\bL = 0$).
The scenes in Figures~\ref{scene} and \ref{scene1}
are related by a radial Lorentz boost.
The advantage of the zero-energy frame
is that it reveals the symmetry between ingoing and outgoing geodesics.
The symmetry is evident in
Figure~\ref{scene}
as a reflection symmetry,
about a horizontal plane passing through the observer,
between the perceived locations of
ingoing (blue) and outgoing (red) surfaces.

The view as the observer approaches the singularity is of particular interest
because,
in the instant that the observer hits the singularity,
null rays in opposite directions along the observer's past lightcone are,
according to observer complementarity,
at the edge of locality,
their future lightcones being at the brink of ceasing to intersect,
Figure~\ref{bhholo_penrose}.
Figure~\ref{zoom}
shows a zoom-in of the view seen
by an observer close to the singularity,
at a radius of $10^{-8}$ geometric units.
The tidal force, which tends to infinity at the singularity,
aberrates the view so that it appears extremely flattened:
emitting surfaces of constant radius appear to be flat planes,
which crowd close to the observer.
In Figure~\ref{zoom},
the observer is on the zero-energy radial geodesic,
but because the tidal force is so overwhelming,
the view is the essentially same regardless
of the radial motion of the observer
(the view looks essentially the same
provided that the observer's geodesic energy satisfies
$| E | \ll \sqrt{- B}$,
which is true as the observer approaches the singularity,
$B \rightarrow - \infty$,
as long as the observer is not accelerating like crazy,
so that $E$ is not diverging).
The view does depend on the observer's angular momentum:
if the observer has some angular momentum,
then the view is aberrated (concentrated) in the direction
of the observer's angular motion.
However,
the view appears flattened into a plane
regardless of the observer's angular motion.

\zoomfig

As the observer approaches the singularity,
the apparent (affine) vertical distance
$\lambda_\obs$
between the observer at radius $r_\obs$
and an emitting surface at radius $r_\emit$
tends to
(the following is the affine distance $\lambda_\obs$
in geometric units
in the zero-energy radial frame,
evaluated
for zero angular momentum photons, $J = 0$)
\begin{equation}
\label{lambdasing}
  \lambda_\obs
  %=
  %{r_\emit - r_\obs \over \sqrt{- B_\obs}}
  \rightarrow
  r_\emit \sqrt{r_\obs / 2}
  \qquad
  \mbox{as $r_\obs \rightarrow 0$}
  \ .
\end{equation}
%The first equality in equation~(\ref{lambdasing})
%is exact in the zero-energy radial frame,
%but it remains a good approximation provided that
%the observer's geodesic energy satisfies
%$| E | \ll \sqrt{- B_\obs}$.
The limit~(\ref{lambdasing})
says that as the observer approaches the singularity,
$r_\obs \rightarrow 0$,
the apparent distance to emitting surfaces
goes to zero as the square root of $r_\obs$.

The arrowed lines converging on the observer in
Figure~\ref{zoom}
represent a sample pair of oppositely directed ingoing and outgoing light rays
along the observer's past lightcone.
Any pair of points on the oppositely directed light rays
(one on the ingoing ray, the other on the outgoing ray)
are at the edge of locality,
meaning that their future lightcones intersect just barely before
the singularity.
Along the past lightcone of an observer hitting the singularity,
only points on oppositely directed ingoing and outgoing light rays
are at the edge of locality.
Points on the lightcone that lie in other than opposing
ingoing and outgoing directions are connected by ``shortcut''
null geodesics, so that their future lightcones intersect
before the singularity.

Figure~\ref{blueshift}
shows the blueshift,
the ratio of observed to emitted photon energy,
observed by an observer close to the singularity,
at a radius of $10^{-8}$ geometric units,
the same as in Figure~\ref{zoom}.
The blueshift is shown for the same five emitting surfaces
shown in Figure~\ref{zoom},
at radii of $4$, $3$, $2.5$, $2$, and $1$ geometric units,
and in addition for emitting surfaces at smaller and larger radii,
$0.01$ and $100$ geometric units.
The blueshift depends on the motion both of the observer and the emitter.
In
Figure~\ref{blueshift},
the observer is taken to be on the zero-energy radial geodesic
($E = 0$, $\bL = 0$),
while the emitters are taken to be radially free-falling from
zero velocity at infinity
($E = 1$, $\bL = 0$),
a natural choice.
While the blueshift depends on the motion of the observer and emitters,
the trend shown by Figure~\ref{blueshift} is generic.
The Figure shows that the blueshift of a point
at a given radius $r_\emit$
is approximately proportional to the affine distance,
but that points at larger emitting radius are more redshifted.

\blueshiftfig

Figure~\ref{blueshift}
shows that as the observer approaches the singularity,
the view above and below the observer
(small affine distances)
becomes highly redshifted,
while the view in the horizontal direction
(large affine distances)
becomes highly blueshifted.
The effect is caused by the enormous tidal force
near the singularity.
Looking up,
the observer feels an upward gravitational force pulling their head off.
Looking down,
the observer feels a downward gravitational force pulling their feet off.
The up and down gravitational forces away from the observer
cause the view above and below to appear gravitationally redshifted.
The same tidal force that pulls the observer apart vertically
also compresses them horizontally.
The horizontal tidal compression
concentrates and blueshifts the view in the horizontal direction.

%DO WE SHOW AND DISCUSS GRAVITY AT HORIZON,
%WHICH RELATES SOMEHOW TO HAWKING TEMPERATURE?

\section{Perceptual distances}
\label{perceptual}

\distfig

The affine distance may be the ``canonical'' measure of distance
along a null geodesic in general relativity,
but
this distance is not directly measurable to an observer
nearing the singularity,
no more than an astronomer can measure the distance to a star
by reaching out and touching it.

\subsection{Binocular distance}

%NEED TO CHECK WEIRD TERM IN $\chi$ BINOCULAR DISTANCE.

A measure of distance
used successfully by apes leaping through trees
is the binocular distance,
also known to astronomers as parallax distance.
The binocular distance is the radius of curvature of
the wavefront of light rays
emitted by the emitter and observed by the observer.

Question for the reader:
if you look at this page through a magnifying glass,
does your binocular vision perceive the text to be
closer or farther away?
Answer at the end of the subsection.

Consider two eyes a small distance $\delta l^\nu$ apart.
The covariant difference
$\delta k^\mu$
in the photon wavevectors
perceived by the two eyes is
\begin{equation}
  \delta k^\mu
  =
  \delta l^\nu {k^\mu}_{;\nu}
\end{equation}
where the colon $;$ represents covariant differentiation.
The eye separation $\delta l^\nu$
is confined to lie on the 3-dimensional future lightcone of the emitter,
and the geodesic equation implies that $\delta k^\mu$ vanishes
for $\delta l^\nu$ parallel to the wavevector $k^\nu$.
Thus without loss of generality the eyes can be taken to lie
in the 2-dimensional subsurface of the emitter lightcone
that is orthogonal to the wavevector
(of course, brain magic allows
binocular vision to work in real animals
regardless of how the eyes are oriented).
More explicitly,
the covariant difference
in photon wavevectors between the two eyes is
\begin{equation}
\label{deltakmu}
  \delta k^\mu
  =
  \delta l^\nu
  \left(
  {\partial k^\mu \over \partial x^\nu}
  +
  \Gamma^\mu_{\kappa\nu} k^\kappa
  \right)
\end{equation}
in which
the first term,
$\delta l^\nu \partial k^\mu / \partial x^\nu$,
represents the change in the photon wavevector
$k^\mu$
between the two eyes,
while the second term,
$\delta l^\nu \Gamma^\mu_{\kappa\nu} k^\kappa$,
represents the effect of parallel-transporting
the photon wavevector from one eye to the other.
The second term corrects
for the difference in spacetime frames of the two eyes
that results from the arbitrariness of coordinates in general relativity.
It is only the covariant difference in photon wavevectors
that is physically measurable.
Evaluating the first term
on the right hand side of equation~(\ref{deltakmu})
involves ray-tracing
from emitter to observer,
while the second term on the right depends only on quantities
local to the observer.

In geometric optics,
the future lightcone of the emitter is a surface of constant phase $\psi$,
the wavefront.
The photon wavevector $k^\mu$
is given by the gradient of the phase,
$k_\mu = \psi_{;\mu} = \partial \psi / \partial x^\mu$
\cite{MTW:1973}.
The null condition
$k^\mu k_\mu = 0$
implies that the photon wavevector $k^\mu$ is orthogonal
to the gradient $k_\mu$ of the phase,
and therefore lies in the lightcone, the surface of constant phase.
The covariant derivative of the photon wavevector
is the curvature of the phase,
a symmetric matrix
in the absence of torsion (as general relativity assumes):
\begin{equation}
  k_{\mu;\nu}
  =
  \psi_{;\mu\nu}
  \ .
\end{equation}
In the locally inertial frame of the observer,
and on the lightcone,
the non-vanishing components of the curvature matrix
constitute a $2 \times 2$ symmetric matrix
in the spatial directions orthogonal to the wavevector.
This is the curvature of the two-dimensional wavefront,
as perceived by the observer.

In flat spacetime (Minkowski space),
the wavefront from a point emitter is spatially spherical,
and the reciprocal of the curvature of the wavefront
equals the affine distance
regardless of how the eyes are oriented about the line of sight.
The fact the binocular distance agrees with the affine distance
in flat space accounts for its successful use by apes.

In curved spacetime,
the curvature matrix has two different eigenvalues,
so the wavefront is ellipsoidal,
yielding two different measures of binocular distance.
The conflicting visual cues would presumably disorient an ape,
just as conflicting visual and motion cues disorient people
inside immersive environments such as airplanes.
Feeling disoriented may be the least of the afflictions
facing an observer nearing the singularity of a black hole.

%In the azimuthal direction, in zero-energy radial frame,
%\begin{equation}
%  \delta k^\mu
%  =
%  \delta l^\nu \Gamma^\mu_{\kappa\nu} k^\kappa
%  =
%  {1 \over r_\obs}
%  \left(
%  \mbox{}
%  -
%  \sqrt{- B}
%  +
%  \cot\theta \sin\chi
%  \right)
%\end{equation}

The top panel of
Figure~\ref{dist}
shows the reciprocal of the polar and azimuthal binocular distances
to the ingoing and outgoing horizons of a Schwarzschild black hole,
as seen by the observer whose view is shown in the top panel of
Figure~\ref{scene}.
The observer is inside the horizon at a radius of $1$ geometric unit,
and is radially infalling on the zero-energy geodesic.
On-axis ($0^\circ$ and $180^\circ$),
the binocular distances agree with the affine distance,
but off-axis the polar and azimuthal binocular distances disagree,
a symptom of the fact that
the wavefront is ellipsoidal rather than spherical.

Answer to the question of whether binocular vision
perceives text seen through a magnifying glass
to be closer or farther away:
farther away.
Try it.

\apefig

\subsection{Trinocular distance}

One must assume that acute inhabitants,
Figure~\ref{ape},
of highly curved spacetime
would evolve three eyes, trinocular vision,
to process the ellipsoidal wavefront
into a best distance,
one that allows them to leap from tree to tree
with the least incidence of death.

The top panel of
Figure~\ref{dist}
shows that the mean of the reciprocal polar and azimuthal binocular distances
agrees well with the affine distance even well off-axis
where the wavefront is quite non-spherical.
%(as indicated by the fact that the individual binocular distances
%differ substantially).
This suggests that
a good strategy for the brains of three-eyed apes
would be to infer a distance from the mean of the
reciprocal binocular distances,
that is,
from the trace of the wavefront curvature matrix.

The agreement of the mean of the reciprocal binocular distances
with the affine distance
can be regarded as a consequence of the Raychaudhuri equation,
which shows that in empty inter-tree space (zero energy-momentum tensor),
the expansion of a bundle of light rays depends only quadratically
on the shear.
Thus the expansion is unaffected to linear order by the shear.
In other words,
the expansion is more or less what it would be in the absence of shear,
provided that the ellipticity of the ray bundle
is less than unity.

The trinocular distance fails as an estimate of affine distance
when the ellipticity of the ray bundle exceeds unity.
The wise ape would learn not to leap at trees
for which the two binocular distances are too discrepant.
This should not pose much of a problem,
since such trees would tend to be farther away,
more likely to be out of leap-range.

As a brief reminder of where the Raychaudhuri and related equations
come from,
a bundle of light rays may be characterized by the four Sachs scalars,
the expansion, vorticity, and complex shear
(see e.g.\ \S9 of \cite{Chandrasekhar:1983}).
The Sachs scalars are equivalent to the
$2 \times 2$ matrix
of tetrad-frame connection coefficents
orthogonal to the wavevector
evaluated in a sequence of locally inertial frames
parallel-transported along the null ray.
The evolution of the Sachs scalars is governed
by the usual equations relating the tetrad-frame connection coefficients
to the Riemann tensor.
In geometric optics, where the wavevector is the gradient of a scalar,
the vorticity vanishes
\cite{MTW:1973}.
If the tidal stress (spin-$2$ component of the Weyl tensor)
along the bundle is non-vanishing,
then shear accumulates along the bundle,
causing the wavefront to become ellipsoidal.
The equation for the expansion,
the Raychaudhuri equation,
depends quadratically on the shear.

\subsection{Angular diameter distance}

Another measure of distance available to the observer
is one known to astronomers as the angular diameter distance.
This distance follows from the perceived angular size
of an object of known proper size.

Whereas the binocular or trinocular distance provides
an unambiguous perceptual measure of distance,
the angular diameter distance does not,
because it requires knowing the actual proper size
of the object being observed,
which the observer may not know.

The observer can deduce an absolute distance
from the fractional difference in angular diameter distances
perceived by two eyes separated along the line-of-sight.
However, this distance is simply the binocular distance,
and so does not provide an additional measure of distance.

The angular diameter distance in any direction is
$\delta l / \delta \alpha$,
where
$\delta l \equiv \sqrt{ \delta l^\nu \delta l_\nu }$
is the proper spatial separation of points at the emitter
transverse to the line-of-sight,
and $\delta \alpha$
is the apparent angle between the points subtended at the observer.
The angular diameter distance
can be calculated by integrating
the equation of geodesic deviation along the null ray,
\begin{equation}
\label{geodesicdeviation}
  {D^2 \delta l^\nu \over D \lambda^2}
  =
  {R_{\kappa \lambda \mu}}^\nu k^\kappa k^\mu \delta l^\lambda
  \ ,
\end{equation}
subject to the initial conditions that
the spatial separation $\delta l^\nu$
lies in the light cone,
is orthogonal to the wavevector,
$\delta l^\nu k_\nu = 0$,
and is initially zero but with non-zero infinitesimal derivative
$D \delta l^\nu / D \lambda$.
The equation of geodesic deviation~(\ref{geodesicdeviation})
is a linear equation for the evolution of deviations $\delta l^\nu$
in the two-dimensional plane transverse to the wavevector.
The transformation matrix
${R_{\kappa \lambda \mu}}^\nu k^\kappa k^\mu$
is symmetric in $\lambda \nu$,
and the solution therefore has two orthogonal eigenvectors.
There are two corresponding angular diameter distances
in orthogonal directions.
Rather than integrate equation~(\ref{geodesicdeviation}) directly,
we prefer to evaluate $\delta l^\nu$ by ray-tracing along the null ray,
\begin{equation}
  \delta l^\nu
  =
  \delta
  \int k^\nu \dd \lambda
  \ ,
\end{equation}
subject to the constraint
that the observed affine distance from observer to emitter is constant,
$\delta \left( E_\obs \int \dd \lambda \right) = 0$:
\begin{eqnarray}
  \delta l^\nu
  &=&
  \int \delta \left( {k^\nu \over E_\obs} \right) E_\obs \, \dd \lambda
\nonumber
\\
  &=&
  \int \delta k^\nu \, \dd \lambda
  -
  \delta \ln E_\obs \int k^\nu \, \dd \lambda
  \ .
\end{eqnarray}

%In the azimuthal direction,
%the perceived angle is $\sin\chi \, \dd \phi$,
%the distance at the emitter is
%$r_\emit \sin\theta \, \dd \phi$,
%and the angular diameter distance is therefore
%\begin{equation}
%  {\dd l \over \sin\chi \, \dd \phi}
%  =
%  {r_\emit \sin\theta \over \sin\chi}
%\end{equation}

The lower panel of
Figure~\ref{dist}
shows the reciprocal of the polar and azimuthal angular diameter distances
to the ingoing and outgoing horizons of the Schwarzschild black hole
as seen by the observer whose view is illustrated at the top of
Figure~\ref{scene}.
Like the two binocular distances,
the two angular diameter distances differ,
but the average of the inverse angular diameter distances
provides a good estimate of the affine distance,
at least as long as the two angular diameter distances
are not too discrepant.

\section{Summary}

The principal purpose of this paper has been to describe
what things look like to an observer who falls inside the horizon
of a Schwarzchild black hole.
Figure~\ref{scene1}
shows a sequence of views of
the apparent location of the ingoing and outgoing horizons of the black hole
from the perspective of an observer who free-falls radially from zero velocity
at infinity.
Figure~\ref{scene}
shows essentially the same thing
from the perspective of an observer who free-falls radially
on the zero-energy geodesic,
where the symmetry between ingoing and outgoing light rays is manifest.

The reader interested in a less abstract portrayal
is invited to view the general relativistically ray-traced visualizations at
\cite{Hamilton:2009}.

The scene in the limit as the observer nears the singularity
is of particular interest
because,
according to observer complementarity,
oppositely directed ingoing and outgoing null rays on the past
lightcone of such an observer are at ``the edge of locality,''
the future lightcones of such rays being on the verge of intersecting.
Figure~\ref{zoom}
shows the view from the perspective of an observer close to the singularity.
Figure~\ref{blueshift}
shows how the various parts of the view appear redshifted or blueshifted.
The diverging tidal field near the singularity
causes the view to appear compressed into a horizontal plane.
The view is blueshifted in the horizontal direction,
but redshifted in all other directions.

An important technical part of this paper
is the problem of assigning a perceived position to any emitter.
An observer in a highly curved spacetime
easily specifies precisely
the perceived angular position of an emitting object,
but,
like an astronomer looking at a star,
has a harder time specifying its distance.
We argue, \S\ref{affine},
that the affine distance provides a definitive measure
of distance along a null ray in curved spacetime.
However, the affine distance is not necessarily directly measurable,
and we therefore discuss, \S\ref{perceptual},
perceptual surrogates for the affine distance.
Binocular distance (parallax distance)
fails in curved spacetime because wavefronts becomes ellipsoidal,
yielding two conflicting measures of binocular distance.
We suggest that acute inhabitants
of highly curved spacetime would evolve three eyes,
Figure~\ref{ape},
to process the ellipsoidal wavefront into a best estimate,
the trinocular distance.
It follows from the Raychaudhuri equation that
the trace of the wavefront curvature matrix yields
a trinocular distance that is a good estimate of the affine distance
provided that the ellipticity of the wavefront is not much greater
than one.

\ack
This work was supported in part by NSF award
AST-0708607.
We thank Wildrose Hamilton for
%drawing
the three-eyed ape,
Figure~\ref{ape}.

\appendix

\section{Affine distance and rigid arms}
\label{rigidarm}

In \S\ref{affine} it was asserted that
if you lived in a highly curved spacetime,
and you put out your rigid arm to touch something,
then the distance that your arm would measure would be the affine distance.

A rigid arm can be defined as one in which
the proper length along any small interval of it is constant,
and equal to its proper length at rest.
If an arm accelerates,
then the only way that the arm can remain perfectly rigid
is for the acceleration at any two points a small interval apart on it
to occur simultaneously in their mutual rest frame,
and to be Rindler-like,
meaning that the acceleration at any point is inversely proportional
to the proper distance to the point along the arm,
starting from some zero point or other.
This kind of acceleration is acausal
(simultaneous over spacelike-separated points).
Thus in relativity,
a body cannot be intrinsically rigid,
because to do so would be to violate causality.
Nevertheless, it is possible to consider an arm that is kept rigid
through superb anticipation and coordination by the brain.

Suppose that you did wield an arm rigidly.
Then the distance it measures would be the affine distance.
This follows from the facts that:
(a) because general relativity is a metric theory,
the proper distance along an interval measured in an accelerating frame
is the same as the proper distance along the interval
measured in a free-fall
%(non-accelerating)
frame
with the same instantaneous velocity;
and
(b) in a free-fall frame,
two points a small interval apart are at rest relative to each other
if and only if
light rays bounced between the points show no Doppler shift.
%IS THAT TRUE FOR RINDLER?

%But, you object,
%if I hang my arm down in a gravitational field,
%won't the end of it appear redshifted?
%And won't the distance that I perceive to the end of the arm
%therefore differ from the affine distance?
%Yes, but that's irrelevant:
%the true distance does not have to be the perceived distance.
%IS THAT RIGHT?

\section*{References}

\bibliographystyle{unsrt}
\bibliography{bh}

\end{document}